\documentclass[twocolumn]{aastex631}

\def\de{\delta}

\begin{document}

\title{\large Transient phenomena from gas erupting neutron stars}


\author{Huiquan Li}
\affil{Yunnan Observatories, Chinese Academy of Sciences, \\
650216 Kunming, China} \affil{Key Laboratory for the Structure and
Evolution of Celestial Objects,
\\ Chinese Academy of Sciences, 650216 Kunming, China}
\affil{Center for Astronomical Mega-Science, Chinese Academy of
Sciences, \\ 100012 Beijing, China}

\author{Jiancheng Wang}
\affil{Yunnan Observatories, Chinese Academy of Sciences, \\
650216 Kunming, China} \affil{Key Laboratory for the Structure and
Evolution of Celestial Objects,
\\ Chinese Academy of Sciences, 650216 Kunming, China}
\affil{Center for Astronomical Mega-Science, Chinese Academy of
Sciences, \\ 100012 Beijing, China}

\begin{abstract}
Starquakes probably occur in rapidly spinning or ultra high field
neutron stars. In this article, we argue that highly
compressed gas containing electron-positron pairs could evaporate
and erupt from inside the neutron star when a crack forms and then
heals during a starquake. Under the influence of the existing
oscillation modes of the star, the crack may have sufficiently large
size and long lifetime. An appropriate amount of gas can erupt into
the magnetosphere with relativistic and nearly uniform velocity,
producing various transient and bursting phenomena.
\end{abstract}



\section{Introduction}
\label{sec:introduction}

Starquake due to crustal failure probably occurs in neutron stars
that spin fast or have strong fields (e.g.,
\citet{2012MNRAS.427.1574L}). In the outer envelopes of these
objects, the shear stress accumulates until the crust cracks.
Starquakes can naturally lead to glitches observed in some pulsars
\citep{1969Natur.223..597R,1991ApJ...382..587R,2008LRR....11...10C}.
In the popular models, the intense crustal fractures cause high
energy flares from magnetars \citep{1995MNRAS.275..255T} and the
small-scale fractures lead to the persistent X-ray emission
\citep{1996ApJ...473..322T}.

However, regarding the fracture model, there remain some doubts. It
is argued by \citep{2003ApJ...595..342J} and further tested in
molecular simulations \citep{2009PhRvL.102s1102H} that no voids or
fractures can occur due to the tremendous isotropic pressure in the
crust. The voids can only exist in extremely short time and heal
very quickly. In magnetars, the strong magnetic fields will impede
the shear motion and so will strongly limit the energy released from
the crack of the crust \citep{2012MNRAS.427.1574L}. Indeed, the
analysis should be applicable to the ideal and simple case. But,
there may still exist mechanisms that can create larger voids or
cracks with a longer lifetime in a realistic neutron star under some
complex circumstances. For example, when the fracture of the crust
is coupled with the existing oscillation modes, the forming crack
may be amplified.

In this work, it is pointed out that the crustal cracks might lead
to gas eruption from the neutron star. Here, we only care about the
size and lifetime of the voids or cracks instead of the intensities
and other properties of the fractures. We keep open on these
quantities of the cracks: they may be very small as anticipated by
the works \citep{2003ApJ...595..342J,2009PhRvL.102s1102H} or somehow
larger due to some unknown mechanisms. Once a crack forms, it will
be quickly filled with gas that comes from the highly compressed
crust or is created locally. The filled gas will later be compressed
and expelled from the crack as the defect heals, producing transient
emission in the magnetosphere. The neutral gas may contain
electro-positron pairs, free neutrons and other plasma of light
elements from inside the crust and the atmosphere. We show that the
energy taken away by the erupted gas can reach the ones of giant
pulses (GPs) and fast radio bursts (FRBs) if the size and lifetime
of the cracks are large enough. This model is different from the
neutron star volcano model \citep{1969Natur.223..486D} where
baryonic lava erupt into the interstellar region from the interior
of the neutron star.

\section{Eruption of gas from neutron stars}
\label{sec:gaseruption}

Once a crack forms as the stress accumulates in the crust, the
pressure in the vacuum crack will suddenly become zero. It will be
quickly filled up with the gas of light, relativistic particles
coming from the highly compressed crust and atmosphere. The
characteristic speed of propagation of the cracks is
\citep{1995MNRAS.275..255T}
\begin{equation}
 c_s=\sqrt{\mu/\rho}\simeq 1.4\times10^8 \textrm{ cm}\cdot
\textrm{s}^{-1},
\end{equation}
where $\mu$ is the shear modulus and $\rho$ is the mass density. The
speed of the relativistic gas is nearly the speed of light. So there
should be enough time for the gas to fill the crack and erupt from
it before the crack heals. We think that the gas contains the
following possible components.

First, the electron-positron pairs. In a young neutron star,
electron-positron pairs can be produced in the presence of high
temperature and magnetic fields. Particularly for a magnetar, the
magnetic field is ultra strong $>10^{15}$ G, much larger than the
Schwinger limit. So the pairs can be created via the one-photon and
the Schwinger processes. On the other hand, the pairs can annihilate
into photons and neutrinos. The latter is one of the key mechanism
that the energy is taken away from neutron star so that a young star
can cool down
\citep{1960PhRvL...5..573C,2001PhR...354....1Y,2008LRR....11...10C}.
In the equilibrium state, the creation rate and the annihilation
rate of the pairs should be equal.

The density of the pairs is highest in the low-density and
high-temperature plasma
\citep{2001PhR...354....1Y,2008LRR....11...10C}. In the crack, the
matter density is zero and there are no degenerate electrons. When a
crack forms, the highly compressed pairs in the crust can drift into
it from nearby regions. Moreover, extra pairs can be created in the
crack simultaneously in the presence of high temperature and
magnetic fields. The pairs should be in the relativistic regime,
which makes them not easy to annihilate. Besides, the strong
magnetic fields should also reduce the annihilation rate with
quantized motion. So it is possible that the crack is filled with
high density pairs.

Second, the gas or liquid in the atmosphere and magnetosphere will
be possibly sucked into the crack as the vacuum crack forms. This
component of plasma is strongly influenced by the magnetic fields.
Only the plasma near the exit of the crack can slide into it along
the magnetic field lines. The plasma contains light elements, like
hydrogen and helium, and electro-positron plasma.

Finally, free neutrons. Deep into the inner crust with increasing
density and pressure, the free neutrons enrich. The free neutrons
are light and neutral. So they can quickly evaporate into the crack
if the pressure is suddenly zero. The neutrons are much heavier than
electrons and they should be minor in the gas.

When the crack forms, it will be loaded with these components of the
gas quickly. When the crack heals, the gas will be expelled out by
the isotropic pressure of the crust. Since the pressure in the crust
is huge, we expect that the erupted gas attains extremely
relativistic velocity. Their velocities should be nearly
uniform\footnote{They should be uniform both in speed and
in direction. The gas should be ejected at almost the same direction,
with a jet-like structure.}
since they are expelled by an almost constant pressure. The Lorentz
factors of the gaseous particles depend
on the energy budget and the component, amount of the particles.

The maximum energy that the gas possibly attains is of the scale:
\begin{equation}
 E=Pv.
\end{equation}
When the density $\rho$ varies from $10^4$ g$\cdot$cm$^{-3}$ to
$10^{14}$ g$\cdot$cm$^{-3}$ in the crust, the average pressure is in
the range $10^{19} \textrm{ dyn}\cdot\textrm{cm}^{-2} <P< 10^{33}
\textrm{ dyn}\cdot\textrm{cm}^{-2}$ (e.g., see
\citet{Haensel:2007yy,2008LRR....11...10C}). Thus, to release large
enough energy, the volume $v$ of the crack is the key in this model.
Another important factor is the lifetime of the crack. A large
enough lifetime is necessary for the gas to have enough time to
escape from the bottom of the crack. More importantly, in a longer
lifetime, the crack is loaded with more gas. More electron-positron
pairs are created and present.

\section{The size and lifetime of a crack}
\label{sec:sizelifetime}

The volume of the erupted gas can be estimated by that of the crack.
We express the latter as
\begin{equation}
 v=ld\de,
\end{equation}
where $l$, $d$ and $\de$ are respectively the length, thickness and
width of the crack. Following the fracture model
\citep{1995MNRAS.275..255T}, the length $l$ can be taken to be about
$1$ km. The thickness should be the scale of that of the crust,
i.e., also $d\sim 1$ km. Thus, the lifetime of the crack should be
larger than $d/c\sim10^{-5}$ s so that the gas at the bottom of the
inner curst has enough time to come out. In the fracture model
\citep{1995MNRAS.275..255T}, the shear displacement is about 100 m
and the crack lifetime is $\sim100 \textrm{ m}/c_s\sim 10^{-4}$ s.
So the gas can escape in time in the model.

However, as analyzed in \citep{2003ApJ...595..342J}, the width of
the crack is several times of the mean distance between ions
($\de\sim 10^{-11}$ cm) since the isotropic pressure in the crust is
huge. So it has an extremely short lifetime $\sim 10^{-19}$ s. Then
the energy taken away by the gas is quite negligible in our model,
following the arguments. For such a short lifetime, the fracture
model \citep{1995MNRAS.275..255T,1996ApJ...473..322T} also does not
work because the shear motion will stop instantly.

But, this might be not necessarily true in some special cases. For
example, the isotropic pressure may be not so strong temporally when
the crack happens on a fast spinning neutron star or a binary
system. The pressure is a total effect of the gravitational and
centrifugal forces on the crust. For rapidly spinning neutron stars,
the centrifugal force is comparable to the gravitational force. At
the moment of fracture, the gravitational force on the broken
tectonic plate can be partially balanced by the centrifugal force.
This may make the crack live longer.

Here, we are more interested in another case. When the fracture of
the crust is coupled with the oscillation modes that are propagating
in the star, the wound of a crack may be amplified. It is known that
neutron stars can sustain various oscillation modes
\citep{1980ApJ...236..899V,1988ApJ...325..725M}. Some of the modes
are left behind after the birth of the star. The energy stored in an
oscillation mode is about $\sim10^{50}$ ergs
\citep{1969ApJ...158....1T,1988ApJ...325..725M}.

As the non-radial oscillation modes that can cause the variation of
density and pressure, the lowest level p-modes have periods of the
scale $\sim$ ms \citep{1969ApJ...158....1T,1988ApJ...325..725M}.
When the oscillating crust cracks suddenly, its dynamics should be
strongly influenced by the varying pressure of the modes. If the
fracture happens to occur near the trough of the dominated wave, the
broken tectonic plates on both sides of the crack will be stretched
in opposite directions. Then the crack may be made wider. It should
have a longer lifetime, close to the period of the mode. The
oscillating modes do not change the total energy and angular
momentum transfer between different parts of the star. So the
glitch behaviour should remain unchanged though the crack is
amplified.

If the width of the amplified crack reaches $\de\sim 1$ cm, then the
energy budget for the erupted gas is
\begin{equation}\label{e:enscale}
 E=10^{40} \frac{P}{10^{30} \textrm{
dyn}\cdot\textrm{cm}^{-2}}\frac{l}{1\textrm{ km}} \frac{d}{1\textrm{
km}}\frac{\de}{1\textrm{ cm}} \textrm{ ergs},
\end{equation}
where we choose the typical pressure of the inner crust
$10^{30}\textrm{ dyn}\cdot\textrm{cm}^{-2}$. This energy is much
less than the oscillation energy of the star.

\section{Transient phenomena}
\label{sec:phenomena}

The free neutrons in the erupted gas are expected to go through the
magnetosphere without interactions since their lifetime is nearly 15
minutes. They may produce observable effect when they reach the
outer medium, where they decay into protons and electrons. But, the
charged particles in the erupted gas will play significant role in
the magnetosphere. The abrupt and short-duration eruption of
electron-positron pairs from the neutron star may be responsible for
the transient phenomena observed from magnetars and fast spinning
neutron stars.

If the charged particles erupt in the open magnetic field line
region with nearly uniform and highly relativistic velocities, they
will slide along the field lines and escape to the far regions.
During the process, the relativistic charges can initiate pair
cascade and create a large amount of pairs additionally. When they
move out to the outer regions where the field becomes weaker and the
cascade mechanism does not effectively work, they produce beamed and
coherent radio emission, which may account for the GPs and FRBs.
The detailed process is quite similar to the ejection and
radiation process of electron-positron pairs discussed in the model
of \citet{2021Innov...200152G}, which is due to the collapse of the
accumulated material accreted from the companion star in the polar
region.

The GPs are very intense and short-duration radio pulses with
respect to average pulses (see
\citet{2007Ap&SS.308..563K,2016JPlPh..82c6302E} for reviews). Our
gas-erupting model may explain the first class of GPs observed from
some young and rapidly spinning neutron stars
\citep{2021MNRAS.501.3900S}. The GPs are 2 to 4 orders stronger than
average pulses and have timescale of nanoseconds to microseconds. So
smaller size or shorter lifetime of the crack than that given in
Eq.\ (\ref{e:enscale}) is needed to explain the GPs. They may be
produced by smaller-scale but more frequent fractures of the crust
\citep{1996ApJ...473..322T}.

FRBs \citep{2007Sci...318..777L} are among the most puzzling
transient phenomena. In a millisecond duration, the characteristic
energy $10^{35}-10^{43}$ ergs can be released during a burst. Their
origin from magnetars is preferred nowadays
\citep{2020Natur.587...54C,2020Natur.587...59B,2020Natur.587...45Z}.
FRBs take similarities to GPs
\citep{2021FrPhy..1624503L} and can be viewed as super-GPs from
young and rapidly spinning pulsars
\citep{2016MNRAS.457..232C,2016ApJ...818...19K,2016MNRAS.462..941L}.
Previously, \cite{2018ApJ...852..140W} revealed evidence
indicating that the repeating FRBs originate from starquake induced
behaviour. Recent work \citep{2021arXiv210401925Y} based on the
fracture model of magnetars is supposed to explain the associated
emission of an FRB at the radio and X-ray band. Here, we provide an
alternative starquake model to interpret the FRB phenomena: they may
be explained as a result of gas eruption from magnetars. Our model
does not require intense fracture of the crust. We only need the
crack to have a large enough volume and lifetime. The typical FRB
energy is reachable as a crack forms with the quantities comparable
to those given in Eq.\ (\ref{e:enscale}).

If the charged particles erupt in the closed field line region, they
will be trapped in this region. Since the curvature of the field
lines is larger in this region, high energy photons can be created
and reprocessed. This may be relevant to the high energy bursts and
flares observed from some magnetars.


\bibliographystyle{aasjournal}
\bibliography{b}

\end{document}